\documentclass[12pt]{iopart}
\usepackage{graphicx}

\newcommand{\PRA}{Phys. Rev. A}
\usepackage{iopams}
\begin{document}
\title[Quantum characterization of superconducting photon
counters]{Quantum characterization of superconducting photon counters}
\author{Giorgio Brida, Luigi Ciavarella, Ivo Pietro Degiovanni, Marco
Genovese, Lapo Lolli, Maria Griselda Mingolla \footnote{also at Dipartimento
di Fisica, Politecnico di Torino, Corso Duca degli Abruzzi 24, Torino I-10129, Italy},
Fabrizio Piacentini, Mauro Rajteri, Emanuele Taralli}
\address{INRIM, Strada delle Cacce 91, Torino I-10135, Italy}
\author{Matteo G. A. Paris}
\address{Dipartimento di Fisica, Universit\`a degli Studi di Milano,
I-20133 Milano, Italy and CNISM, Udr Milano, I-20133 Milan, Italy}
\begin{abstract}
We address the quantum characterization of photon counters based on
transition-edge sensors (TESs) and present the first experimental
tomography of the positive operator-valued measure (POVM) of a TES. We
provide the reliable tomographic reconstruction of the POVM elements
up to $11$ detected photons and $M=100$ incoming photons, demonstrating
that it is a linear detector.
\end{abstract}
\maketitle
\section{Introduction}
The possibility of discriminating the number of impinging photons on a
detector is a fundamental tool in many different fields of optical
science and technology \cite{had09}, including nanopositioning and the
redefinition of candela unit in quantum metrology \cite{pol09,zwi10},
foundations of quantum mechanics \cite{gen05}, quantum imaging
\cite{bri10} and quantum information \cite{lan09,obr09,gis07,ben10}, e.g
for communication and cryptography. As a matter of fact, conventional
single-photon detectors can only distinguish between zero and one (or
more) detected photons, with photon number resolution that can be
obtained by spatially \cite{multiSpatial} or temporally
\cite{multiTemporal} multiplexing this kind of on/off detectors.
\par
Genuine Photon Number Resolving (PNR) detectors needs a process
intrinsically able to produce a pulse proportional to the number of
absorbed photons. In fact, detectors with PNR capability are few, e.g.
photo-multiplier tubes \cite{burle}, hybrid photo-detectors \cite{NIST}
and quantum-dot field-effect transistors \cite{gan07}. At the moment,
the most promising genuine PNR detectors are the visible light photon
counters \cite{yam99} and Transition Edge Sensors (TESs)
\cite{fuk11,pre09,cab08,lit08,ros05,bandler06}, i.e. microcalorimeters
based on a superconducting thin film working as a very sensitive
thermometer \cite{irw05}.
\par
For a practical application of these detectors it is crucial to achieve
their precise characterization
\cite{lss99,jar01,dem03,mit03,dar04,zam05,lob08,rah11}.  In particular,
it is generally assumed that TESs are linear photon counters, with a
detection process corresponding to a binomial convolution.  It is also
expected that dark counts are not present in TESs. Taken together, these
assumptions allow one to characterize a TES by a single number assessing
the quantum efficiency of the detector, i.e. the probability $0\leq \eta
\leq 1$ that a photon impinging onto the detector is actually revealed.
In this paper, we present the first experimental reconstruction of the
POVM describing the operation of a TES and, in turn, the first
demonstration of the linearity. In section 2 we illustrate the method used for POVM
reconstruction, while in section 3 we describe the
experimental implementation. In section 4 we discuss the results and
close the paper with some concluding remarks.
\section{POVM reconstruction technique}
As TESs are microcalorimeters, they are intrinsically phase insensitive
detectors. In the following we thus assume that the elements of the
positive operator-value measurement (POVM) $\{\Pi_n\}$ are diagonal
operators in the Fock basis, i.e.  \begin{equation} \Pi_n=\sum_m
\Pi_{nm} |m\rangle \langle m|, \end{equation} with completeness relation
$\sum_n \Pi_n = \mathbf{I}$. Matrix elements $\Pi_{nm}=\langle m| \Pi_n|
m\rangle$ describe the detector response to $m$ incoming photons, i.e.
the probability of detecting $n$ photons with $m$ photons at the
input\footnote{This corresponds to consider our TES as a \emph{grey box}
(instead of a black box), on the basis of this solid physics assumption,
i.e. the fact that they are microbolometers. On the other hand, trying
to find an experimental evidence of this phase-insensitiveness
assumption is pointless, as there is not a phase reference (e.g. from
the TES itself) to modify the phase of our probe states with respect to
it.}. A reconstruction scheme for $\Pi_{nm}$, i.e. a tomography of the
POVM, provides the characterization of the detector at the quantum
level. In order to achieve the tomography of the TES POVM, we exploit an
effective and statistically reliable technique \cite{h1,lun09,h2} based
on recording the detector response for a known and suitably chosen set
of input states, e.g. an ensemble of coherent signals providing a sample
of the Husimi Q-function of the elements of the POVM.  \par Let us
consider a set of $K$ coherent states of different amplitudes
$|\alpha_j\rangle$, $j=1,...,K$. The probability of obtaining the
outcome $n$ from the TES, i.e. of detecting $n$ photons, with the $j$-th
state as input is given by
\begin{equation}
\label{eq:stat_model}
p_{nj} =\hbox{Tr}[|\alpha_j\rangle\langle\alpha_j| \Pi_n] =\sum_m \Pi_{nm}\, q_{mj}
\end{equation}
where $q_{mj}=\exp(-\mu_j)\mu_j^m/m!$ is the ideal photon statistics of
the coherent state $|\alpha_j\rangle$, $\mu _j = |\alpha_j|^2$ being the
average number of photons. In order to reconstruct the matrix elements
$\Pi_{nm}$, we sample the probabilities $p_{nj}$ and invert the
statistical model composed by the set of Eqs. (\ref{eq:stat_model}).
Since the Fock space is infinite dimensional, this estimation problem
contains, in principle, an infinite number of unknowns.
\par
A suitable truncation at a certain dimension $M$ should be performed, with
the constraint that the probability of having $m\geq M$ photons in the
states $|\alpha_j\rangle$ is not too large. In other words, given the
set of probing coherent states, we have a little amount of data for
the entries with $m \geq M$ and we cannot
investigate the performances of the detector above the corresponding
energy regimes.
\par
The distributions $p_{nj}$ in Eq. (\ref{eq:stat_model}) provide a sample
of the Q-functions $\langle\alpha_j| \Pi_n|\alpha_j\rangle$ of the POVM
elements, and any reconstruction scheme for the $\Pi_{nm}$ basically
amounts to recover the Fock representation of the $\Pi_n$'s from their
phase space Q-representation. In general, this cannot be done exactly
due to singularity of the antinormal ordering of Fock number projectors
$|n\rangle\langle n|$ \cite{bal83}. On the other hand, upon exploiting
the truncation described above, we deal with POVM elements expressed as
finite mixture of Fock states, which are amenable to reconstruction
\cite{qsm,dmr}.  The statistical model in (\ref{eq:stat_model}) may be
solved using maximum likelihood (ML) methods or a suitable approximation
of ML. We found that reliable results are obtained already with a least
squares fit, i.e we have effectively estimated $\Pi_{nm}$ by minimization
of a regularized version of the square difference
$\sum_{nj} (\sum_{m=0}^{M-1} q_{mj}\, \Pi_{nm} - p_{nj})^2$
where the physical constraints of smoothness is implemented by exploiting
a convex, quadratic and device-independent function \cite{lun09}. We
also force normalization $\sum_{n=0}^{11} \Pi_{nm}=1$, $\forall m$,
where the last POVM element is defined as $\Pi_{11}=1-\sum_{n=0}^{10}\Pi_{n})$.
\section{Experiment}
The TES we have characterized is composed by a $\sim$ \mbox {90 nm}
thick Ti/Au film \cite{por08,tar07}, fabricated by e-beam deposition on
silicon nitride substrates. The effective sensitive area, obtained by
lithography and chemical etching, is $20\times20$ $\mu$m. The
superconducting wirings of Al, with thicknesses between 100 nm and 150
nm, have been defined by a lift-off technique combined with RF
sputtering of the superconducting films. Upon varying the top Ti film
thickness, the critical temperatures of these TESs can range between 90
mK and 130 mK, showing a sharp transition (1-2 mK).
\par
The characterization of TES has been carried out in a dilution
refrigerator with a base temperature of 30 mK.  Furthermore, the
detector is voltage biased, in order to take advantage of the negative
Electro-Thermal Feedback, providing the possibility to obtain a
self-regulation of the bias point without a fine temperature control and
reducing the detector response time. The read-out operations on our TES
is performed with a DC-SQUID current sensor \cite{dru07}.  Using room
temperature SQUID electronics, we bias our device and read out the
current response. Finally, the SQUID output is addressed to a LeCroy 400
MHz oscilloscope, performing the data acquisition, first elaboration and
storage. In our experiment, we have illuminated the TES with a
power-stabilized fiber coupled pulsed laser at $\lambda=1570$ nm (with a
pulse duration of 37 ns and a repetition rate of 9 kHz), whose pulse is
also used to trigger the data acquisition for a temporal window of 100
ns.  The laser pulse energy $(365\pm 2)$ pJ is measured by a calibrated
power meter, and then attenuated to photon counting regime exploiting
two fiber coupled calibrated attenuators in cascade. The attenuated
laser pulses are then sent to the TES detection surface by a single mode
optical fiber.  The set of coherent states needed to perform the POVM
reconstruction has been generated by lowering the initial laser pulse
energy from an initial attenuation of 63.5 dB {(corresponding to
an average of 130 photons per pulse)}, to 76.5 dB {(mean photon
number per pulse: 6.5)}, to obtain 20 different states $|\alpha_j
\rangle =|\sqrt{\tau_j} \alpha\rangle$ where $\tau_j$ is the channel
transmissivity, $j=1, ..., 20$.
\begin{figure}[h]
\begin{center}
\includegraphics[width=\columnwidth]{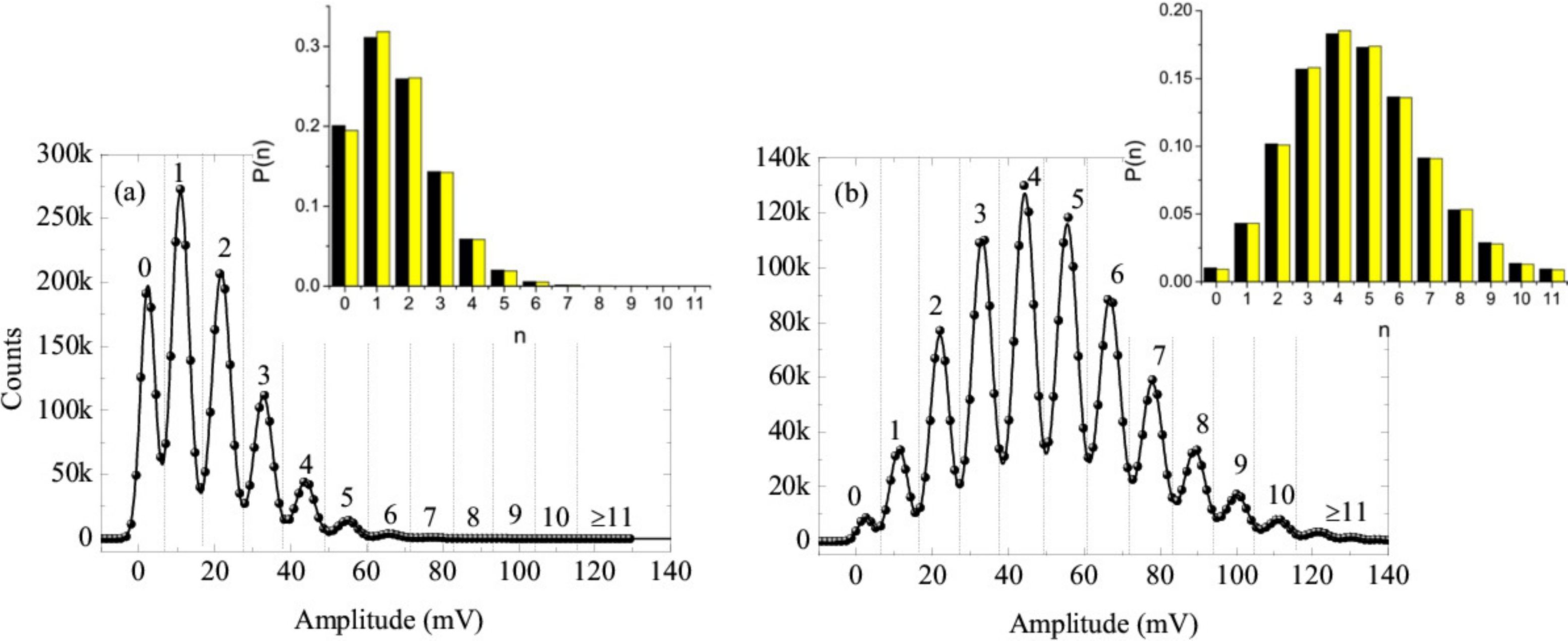}
\caption{{Dots represent the TES counts for two different values of
$|\alpha_j \rangle$: each point corresponds to a binning of an amplitude
interval of $1.3$ mV. Solid lines are the Gaussian fits on the
experimental data, while the dotted vertical lines are the thresholds.
Figure (a) is obtained with a coherent state characterized by a mean
photon number per pulse $\mu=31$, while for figure (b) the state used
had $\mu=87$. The insets of both figures compare the experimental
probability distribution (black bars), obtained from measurements binned
according to the drawn thresholds, with the corresponding Poisson
distributions of mean value $ \eta\mu$ (with $\eta= 5.1 \%$) (yellow
bars): as evident from the plots, the experimental results are in
remarkable agreement with the theoretical predictions, showing
respectively a fidelity of $99.994\%$ and $99.997\%$ .} \label{f1_TES}}
\end{center}
\end{figure}
\par
We work at fixed wavelength $\lambda=1570$ nm and thus, in ideal
conditions, we would expect a discrete energy distribution with outcomes
separated by a minimum energy gap $\Delta E=\frac{hc}{\lambda}$.
Experimentally, we observe a distribution with several peaks, whose
variances represent the energy resolution of the whole detection device.
In a first calibration run, after a binning on the oscilloscope
channels, we fit the data with a sum of independent Gaussian functions
(Fig. \ref{f1_TES} shows that the fitting functions are in excellent
agreement with experimental data); the first peak on the left is the
``0-peak", corresponding to no photon detection.  These fits allowed us
to fix the amplitude thresholds (located close to the local minima)
corresponding to $n$ detected photons: this way, the histogram of counts
is obtained just binning on the intervals identified by these
thresholds. The distributions $p_{nj}$ are finally evaluated upon
normalizing the histogram bars to the total number of events for the
given state \footnote{Remarkably, the reconstructions obtained by
binning data using thresholds are almost indistinguishable from the ones
obtained by evaluating the number of events in the $n-th$ peak by
integrating the corresponding Gaussian of the fit reported in Fig.
\ref{f1_TES}.}. This threshold-based counts binning may introduce some
bias or fluctuations since the tails of the $n$-th Gaussian peak fall
out of the $n$ counts interval. On the other hand, the effects in
neighbouring peaks compensate each other and, overall, do not affect the
tomographic reconstruction.
\section{Results}
\begin{figure}[h!]
\begin{center}
\includegraphics[width=0.75\columnwidth]{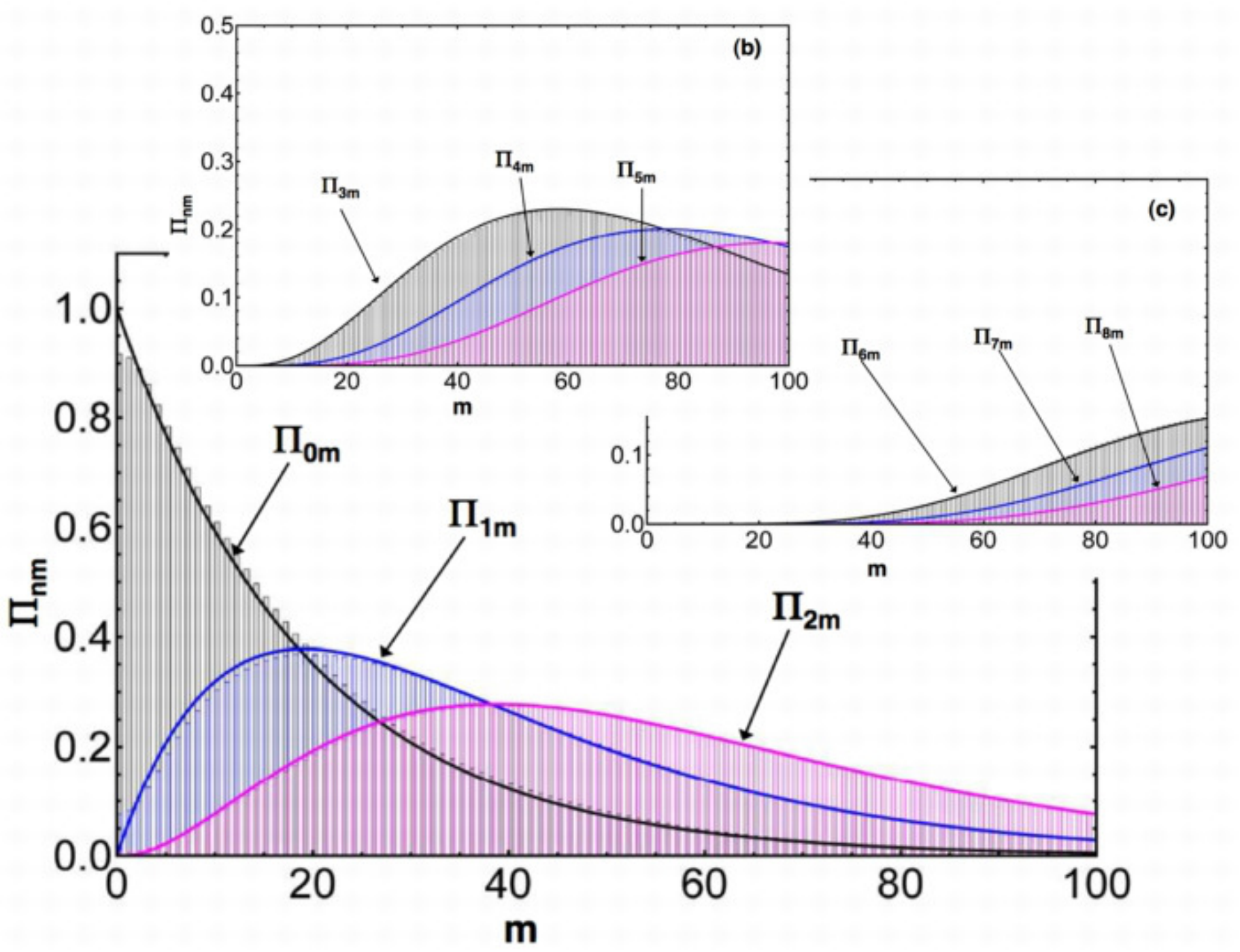}
\caption{(Color Online) Reconstructed POVM of our TES photon counting
systems. Bars represent the matrix elements $\Pi_{nm}$ as a function of
$m=0,100$ for $n=0,1,2$ (main plot), $n=3,4,5$ (b), $n=6,7,8$ (c).
Continuous lines represent the POVM elements of a linear photon counter
with quantum efficiency $\eta=5.10\%$.} \label{f2_POVM}
\end{center}
\end{figure}

The POVM of our TES detection system has been reconstructed up to
$M=140$ incoming photons and considering $N=12$ POVM elements $\Pi_n$,
$n=0,...,N-1$, with $\Pi_{N-1}= \mathtt{1}- \sum_{n=0}^{N-2}\Pi_{n} $
describing the probability operator for the detection of more than $N-2$
photons. In Fig. \ref{f2_POVM} we show the matrix elements $\Pi_{nm}$ of
the first 9 POVM operators ($n=0,..,8$), for $0\leq m\leq 100$. The bars
represent the reconstructed $\Pi_{nm}$, while the solid lines denote the
matrix elements of a linear detector. In fact, as mentioned above, the
POVM of a linear photon counter can be expressed as a binomial
distribution
\begin{equation}
\Pi_n=\sum_{m=n}^{\infty} B_{nm} |m\rangle\langle m|
\end{equation}
of the the ideal photon number spectral measure with $B_{nm}
=\left(\begin{array}{c} m \\ n \end{array}\right) \eta^n (1-\eta)^{m-n}
$, where $\eta$ is the quantum efficiency of the detector.  In order to
compare the POVM elements of the linear detector, i.e. $B_{nm}$, with
the reconstructed POVM elements $\Pi_{nm}$ we have first to estimate the
value of the quantum efficiency $\eta$.
\par
This can be done on the sole basis of the experimental data using ML
estimation, i.e. we average the values of $\eta$ which maximize the
log-likelihood functions
\begin{equation}
L_j=\sum_{n} N_{nj} \log\left(\sum_m B_{nm} q_{mj}\right)
\end{equation}
where $N_{nj}$ is the number of $n$-count events obtained with the
$j$-th input state $|\sqrt{\tau_j}\alpha\rangle$.  The overall procedure
leads to an estimated value of the quantum efficiency $\eta=(5.10 \pm
0.04) \%$, where the uncertainty accounts for the statistical
fluctuations (for each signal probe we estimated the value of $\eta$,
and then we averaged over the ensemble).
\par
\begin{figure}[h!]
\begin{center}
\includegraphics[width=0.6\columnwidth,angle=270]{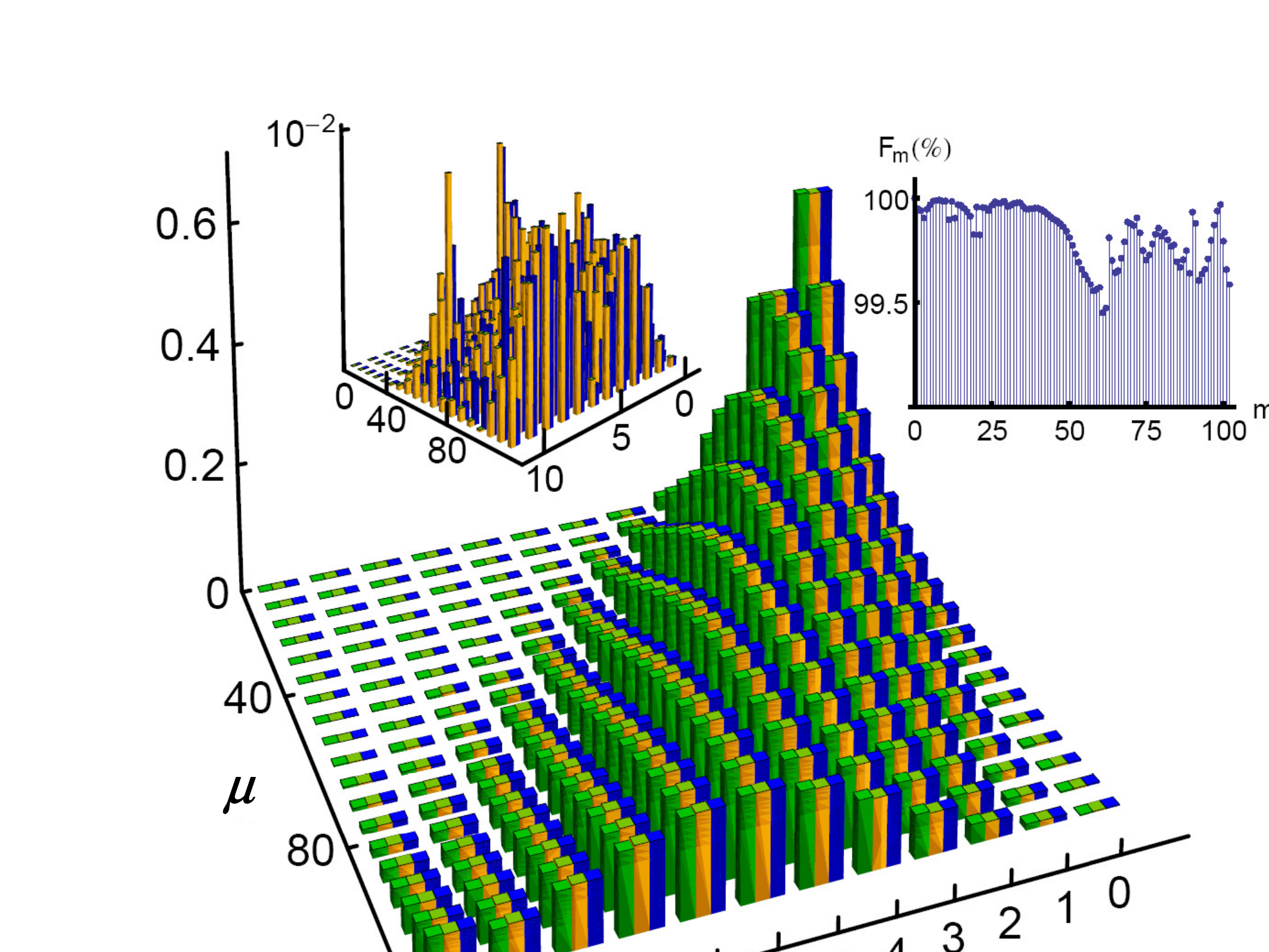}
\vspace{15mm}
\caption{(Color Online) Comparison of the measured distributions
$p_{nj}$ (green bars, on the left of each group) of the coherent states
$|\alpha_j\rangle$ used for POVM reconstruction with those obtained
using the reconstructed POVM elements $r_{nj}$ (yellow central bars).
and with those obtained under the linearity hypothesis $l_{nj}$ (blue
right bars) The left inset shows the absolute differences
$|p_{nj}-r_{nj}|$ (yellow left bars) and $|p_{nj}-l_{nj}|$ (blue right
bars).  The right inset shows the fidelity $F_m$ between the
reconstructed POVM elements at fixed $m$ and those of a linear photon
counter with quantum efficiency $\eta=5.10\%$. }\label{hus} \end{center}
\end{figure}

As it is apparent from Fig. \ref{f2_POVM}, we have an excellent
agreement between the reconstructed POVM and the linear one with the
estimated quantum efficiency. In particular, the elements of the
POVM are reliably reconstructed for $m \leq 100$, whereas for higher
values of $m$ the quality of the reconstructions degrades.
In the regime $m \leq 100$ the fidelity $F_m=\sum_n \sqrt{\Pi_{n m} B_{n
m}}$ is larger than 0.99 (see the right inset of Fig. \ref{hus}), while
it degrades to 0.95 for $100 \leq m \leq 140$. In order to investigate
the effects of experimental uncertainties, we performed a sensitivity
analysis taking into account the uncertainties on the energy of the
input state and on the attenuators, obtaining fidelities always greater
than $98.35\%$ for the $12$ entries.  In order to further confirm the
linearity hypothesis, as well as to assess the reliability of the
reconstruction, we have compared the measured distributions $p_{nj}$
with those obtained for a linear detector, i.e.
\begin{equation}
l_{nj}= \eta^n \exp(-\eta \mu_j) \mu_j^n/n!
\end{equation}
 and with those obtained using the reconstructed
POVM elements, i.e.
\begin{equation}
r_{nj}=\sum_{m=n}^{M} \Pi_{nm} q_{mj} .
\end{equation}
In Fig. \ref{hus} we report the three distributions for the whole set of
probing coherent states, whereas in the left inset we show the (absolute)
differences $|p_{nj}-l_{nj}|$ and $|p_{nj}-r_{nj}|$ between those
distributions and the measured ones.
\par
As it is apparent from the plots,
we have an excellent agreement between the different determinations of
the distributions. This confirms the linear behavior of the detector,
and proves that the reconstructed POVM provides a reliable description
of the detection process.
We have also modified the detection model to take into account the
possible presence of dark counts. In this case, upon assuming a
Poissonian background, the matrix elements of the POVM are given by
$\Pi_{nm}= \exp(-\gamma) \sum_j\gamma^j/j!\,B_{(n-j)m}$ and we have
developed a ML procedure to estimate both the quantum efficiency $\eta$
and the mean number of dark counts per pulse $\gamma$. We found that the
value for $\eta$ is statistically indistiguishable from the one obtained
with the linear-detector model, whereas the estimated dark counts per
pulse are  $\gamma=(-0.03
\pm 0.04)$, in excellent agreement with the direct measurement performed
on our TES detector using the same fitting technique discussed above, providing a substantially negligible dark count level 
$\gamma=(1.4\pm0.6)\times10^{-6}$. The same conclusion is obtained for
any other model, e.g. super-Poissonian, of the background.
\par
In conclusion, we have performed the first tomographic reconstruction of
the POVM describing a TES photon detector. Our results clearly validate
the description of TES detectors as linear photon counters and, together with the precise estimation of the
quantum efficiency, pave the way for practical applications of TES
photon counters in quantum technology.
\section*{Acknowledgements}
We acknowledge the support of the EC FP7 program ERA-NET Plus, under
Grant Agreement No. 217257.  MGAP thanks Stefano Olivares for useful
discussions.

\end{document}